\documentclass[%
 reprint,
superscriptaddress,
 amsmath,amssymb,
 aps,
]{revtex4-2}

\usepackage[utf8]{inputenc}

\usepackage{graphicx}
\usepackage{dcolumn}
\usepackage{bm}
\usepackage{color}

\begin{document}

\title{Resonant enhancement of particle emission from  a parametrically driven  condensate in a one-dimensional lattice}
\author{L. Q. Lai}
\affiliation{School of Physics and Electronics, Hunan University, Changsha 410082, China}
\affiliation{Laboratory of Atomic and Solid State Physics, Cornell University, Ithaca, New York 14853, USA}

\author{Y. B. Yu}
\affiliation{School of Physics and Electronics, Hunan University, Changsha 410082, China}

\author{Erich J. Mueller}
\email{em256@cornell.edu}
\affiliation{Laboratory of Atomic and Solid State Physics, Cornell University, Ithaca, New York 14853, USA}

\date{\today}

\begin{abstract}

Motivated by recent experiments, we investigate particle emission from a Bose-Einstein condensate in a one-dimensional lattice, where the interaction strength is periodically modulated.
The modulated interactions parametrically excite a collective mode, leading to density oscillations.  These collective oscillations in turn drive particle emission.  This multistep process amplifies the drive, producing larger particle jets.  We find that the amplitude dependence of the emission rate has a characteristic threshold behavior, as seen in experiments.

\end{abstract}

\maketitle

\section{Introduction}

Pattern formation is one of the most striking phenomena of
nonequilibrium dynamical systems \cite{pattern}. Examples can be found in contexts ranging from fluid dynamics and nonlinear optics \cite{fluid,optics} to biochemistry and the early universe \cite{bio,universe}. In the quantum regime the physics is even richer \cite{quantum1,quantum2,quantum3,quantum4,quantum5,quantum6,quantum7,quantum8}, with the possibility of new forms of ordering and  superpositions of macroscopically distinguishable patterns. Here we model cold atom experiments where periodically modulating the interaction strength leads to particle “jets” \cite{chin1,chin2,chin3,chin4}, focusing on the role of collective modes.

In a previous work we analyzed a minimal model of these experiments, which demonstrated how the modulated interactions lead to particle emission \cite{lai}. That previous model consists of a semi-infinite one-dimensional (1D) lattice where a deep local potential confines a condensate to the boundary site.  When the interactions are modulated at the appropriate frequencies, atoms can be excited from the condensate into unbounded modes, allowing them to escape.
While that model provided significant insights, it did not contain enough degrees of freedom to capture the role of collective modes. Unlike the experiments, particle  emission was found for arbitrarily weak drive strength, and it was incapable of explaining the observed density modulations. To overcome these deficiencies, here we introduce a slightly more sophisticated model, in which the trap consists of two sites. The modulated interactions can then parametrically excite a ``sloshing mode," which in turn drives particle emission.  This collective mode provides a resonant enhancement, allowing a weak modulation to produce a jet containing a macroscopic number of particles.  Evidence of these collective modes were found in the experiment \cite{chin2}.
Our model explains the large density modulations which accompany the particle emission, as well as the amplitude dependence of the emission rate.  In particular we give a physical picture of the threshold behavior seen in Ref.~\cite{chin1}.





There has been a number of theoretical works which explained various aspects of the experiments \cite{chin3,chin4,yan,zhai,holland}.
Our model differs from previous ones in its simplicity.  We are able to investigate all of the relevant physics in a transparent way, and discuss the conditions for resonantly exciting the collective mode, the mode-matching conditions with the environment, and the dependence on drive amplitude.
{{}
Related physics
has also been explored in other contexts \cite{riboli,lee,whitlock,haroutyunyan,salasnich,gati,topfer}.
}

In Sec.~\ref{sec:model} we describe our model and the relevant equations of motion. In Sec.~\ref{sec:perturb} we perturbatively analyze this model with respect to the symmetric mode and the antisymmetric mode. In Sec.~\ref{sec:numerics} we parametrically drive the antisymmetric mode and compare the results to a numerical solution. We provide a summary in Sec.~\ref{sec:summary}.

\section{Model}\label{sec:model}

\begin{figure}[tbp]
\includegraphics[width=\columnwidth]{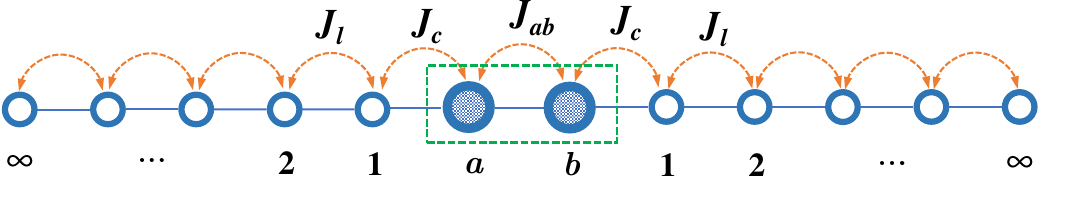}
\caption{(color online) Schematic of the 1D infinite lattice. A local trapping potential $V$ is applied to the dashed box, containing sites labeled $a$ and $b$.  The sites on the leads are labeled by non-zero integers.}
\label{latticepic}
\end{figure}

We consider a 1D infinite lattice, as depicted in Fig.~\ref{latticepic}.  The two central sites, labeled $a$ and $b$, represent a trap of depth $V$ which confines a Bose-Einstein condensate. Atoms in these sites can tunnel back and forth, with amplitude $J_{ab}$. If an atom has sufficient energy, it could also escape from the trap, by hopping onto one of the two leads, with sites labeled $1,2,\cdots$.  The coupling between the trap and the lead has amplitude $J_c$, while $J_l$ quantifies the coupling between nearest-neighboring sites in each lead.  Since the particle density is small outside of the trap, we can neglect the interactions there, and write the Hamiltonian as
\begin{eqnarray}
\hat{H} &=&V\left( \hat{a}_{0}^{\dag }\hat{a}_{0}+\hat{b}_{0}^{\dag }\hat{b}_{0}\right) \nonumber \\
&&+\frac{1}{2}\left[ U+g\left( t\right) \right] \left( \hat{a}_{0}^{\dag
}\hat{a}_{0}^{\dag}\hat{a}_{0}\hat{a}_{0}+\hat{b}_{0}^{\dag }\hat{b}_{0}^{\dag }\hat{b}_{0}\hat{b}_{0}\right)  \nonumber \\
&&-J_{ab}\left( \hat{a}_{0}^{\dag }\hat{b}_{0}+\hat{b}_{0}^{\dag }\hat{a}_{0}\right)  \nonumber  \\
&&-J_{c}\left( \hat{a}_{0}^{\dag }\hat{a}_{1}+\hat{a}_{1}^{\dag }\hat{a}_{0}+\hat{b}_{0}^{\dag
}\hat{b}_{1}+\hat{b}_{1}^{\dag }\hat{b}_{0}\right)  \nonumber \\
&&-J_{l}\sum_{j=1}^{\infty }\left( \hat{a}_{j+1}^{\dag }\hat{a}_{j}+\hat{a}_{j}^{\dag
}\hat{a}_{j+1}+\hat{b}_{j+1}^{\dag }\hat{b}_{j}+\hat{b}_{j}^{\dag }\hat{b}_{j+1}\right),
\end{eqnarray}
where
$\hat{a}_{j}^{\dag}      (\hat{a}_{j})$ and $\hat{b}_{j}^{\dag} (\hat{b}_{j})$ are creation (annihilation) operators on the $j$th site to the left or right; $\hat{a}_0$ and $\hat{b}_0$ correspond to the trapped sites. The time-dependent pairwise interactions are characterized by a constant term $U$ and a sinusoidally oscillating term $g(t)=g \sin(\omega t)\theta(t)$, where $\theta(t)$ is the step function. In the experiments, the dc component of the interactions is generally small \cite{chin1,chin2,chin3,chin4}.  Thus, to simplify the analysis, we take the limit $U=0$.  In our previous work, where the geometry was somewhat simpler, we extensively studied $U\neq0$ and found that finite $U$ played an insignificant role in the physics, while making the analysis much more complicated.

Experimentally, this inhomogeneous lattice, with a trap and barriers, could be implemented in a quantum-gas microscope \cite{kuhr} or in a hybrid system involving an optical lattice and optical microtraps \cite{tweezer}.  Time- and space-dependent interactions are  routinely implemented through  magnetic field driven Feshbach resonances \cite{chin}.  We emphasize, however, that the value of this model is not in describing a particular experiment, but in providing a simple context to explore the physics.  The  motivating experiments \cite{chin1,chin2,chin3,chin4} were performed in the continuum, and faithfully modeling them requires a more complicated and hence less transparent formalism.

The trap and the jets contain a macroscopic number of particles, thus it is reasonable to replace
the operators  with their expectation values $a_j=\langle\hat a_j\rangle$ and $b_j=\langle\hat b_j\rangle$. Physically $\vert a_j \vert^2$ and $\vert b_j \vert^2$ represent the number of particles on site $j$ in each lead.
Using units where $\hbar=1$, the expectation value of the Heisenberg equations of motion read
\begin{eqnarray}
i\partial _{t}a_{0} &=&\langle[ \hat{a}_{0},\hat{H}]\rangle \nonumber \\ \label{a0}
&=&Va_{0}+g\sin\left(\omega
t\right)  \left\vert a_{0}\right\vert ^{2}a_{0}-J_{ab}b_{0}-J_{c}a_{1}, \\\label{a1}
i\partial _{t}a_{1} &=&\langle[ \hat{a}_{1},\hat{H}]\rangle =-J_{c}a_{0}-J_{l}a_{2}, \\\label{aj}
i\partial _{t}a_{j} &=&\langle[ \hat{a}_{j},\hat{H}]\rangle =-J_{l}\left(
a_{j-1}+a_{j+1}\right). \label{eomj}
\end{eqnarray}
Similar equations hold for $b_j$. In equilibrium where $g=0$, we find a stationary state of the form $a_0=b_0=\alpha e^{-i \nu t}$,
$a_{j\geq 1}=b_{j\geq 1}=\alpha e^{-i \nu t} e^{-\kappa_1} e^{-\kappa (j-1)}$.  Straightforward algebra gives
$\nu=\frac{(J_c^2-2J_l^2)(V-J_{ab})\pm J_c^2 \sqrt{(V-J_{ab})^2+4J_c^2-4J_l^2}}{2(J_c^2-J_l^2)}$,
$\cosh\kappa=\frac{\nu}{-2 J_l}$,
and
$\kappa_{1}=-{\rm ln}(\frac{-J_c}{\nu+J_l e^{-\kappa}})$.



As in our previous work, we can use Green's function techniques to eliminate the leads, solving  Eqs. (\ref{a1})  and (\ref{aj}) to write $a_1$ as a function of $a_0$.  This results in  a set of nonlinear integrodifferential equations for the order parameters in the trap,
\begin{eqnarray}
i\partial _{t}a_{0} &=&Va_{0}-J_{ab}b_{0}+J_{c}^2 \int^t G_{11}(t-\tau)a_0(\tau)d\tau \nonumber \\
&&+g\sin(\omega t) |a_0|^2 a_0, \label{master1} \\
i\partial _{t}b_{0} &=&Vb_{0}-J_{ab}a_{0}+J_{c}^2 \int^t G_{11}(t-\tau)b_0(\tau)d\tau \nonumber \\
&&+g\sin(\omega t) |b_0|^2 b_0, \label{master2}
\end{eqnarray}
where $G_{j1}$ is the time-domain Green's function
\begin{eqnarray}
G_{j1}(t)= i^{j-2} \frac{j J_j(2 J_l t)}{J_l t} \theta(t)
\end{eqnarray}
with $J_n(z)$ the Bessel function of the first kind. We can numerically solve these equations, using the techniques discussed in Ref.~\cite{lai}.  Equivalently, we can truncate the leads and directly solve Eqs.~(\ref{a0})--(\ref{aj}).  Our key results in the following come from perturbatively solving Eqs.~(\ref{master1}) and (\ref{master2}).

\section{Perturbative Analysis}\label{sec:perturb}
In the absence of the drive ($g=0$) the equations of motion are linear.  The resulting spectrum has two discrete peaks, and a continuum. The peaks represent the symmetric and antisymmetric modes in the trap, and the continuum corresponds to the modes in the lead.  In the limit where $J_c$ is negligible the discrete modes give frequencies $\nu_s=V-J_{ab}$ and $\nu_a=V+J_{ab}$.  The continuum corresponds to $-2 J_l<\nu<2 J_l$.

To observe resonantly enhanced emission, we need to be in the regime where the symmetric mode is stable (outside the continuum) but the antisymmetric mode is unstable (inside the continuum), i.e.,
\begin{equation}\label{constraint}
|V-J_{ab}|>2 J_l,\quad |V+J_{ab}|<2 J_l.
\end{equation}
Under these circumstances a large-amplitude antisymmetric mode can decay into particle jets.  We will parametrically excite this antisymmetric mode by modulating the interaction strength, taking $\omega=2(\nu_a-\nu_s)=4J_{ab}$. In our previous work
we had one fewer constraint, only requiring that the trapped mode was stable and that a multiple of the drive frequency connected the bound state to the continuum.

To model the emission process, we use the method of multiple scales, writing
\begin{eqnarray}
a_0&=&e^{-i \nu_s t} \psi(t) + e^{-i \nu_a t} \phi(t),\\
b_0&=&e^{-i \nu_s t} \psi(t) - e^{-i \nu_a t} \phi(t),
\end{eqnarray}
where $\psi(t)$ and $\phi(t)$ are the slowly varying amplitudes of the symmetric and antisymmetric modes, respectively. We substitute this ansatz into Eqs.~(\ref{master1}) and (\ref{master2}), and use that for any slowly varying function $f(t)$,
\begin{eqnarray}
\int^t G_{11}(t-\tau)e^{-i\nu \tau} f(\tau)d\tau
&\approx&
 f(t)e^{-i\nu t} G_{11}(\nu).
\end{eqnarray}
Discarding the rapidly oscillating terms in the resulting expression yields
\begin{eqnarray}\label{psi}
i\partial_t\psi &=& J_c^2 G_{11}(\nu_s) \psi +\frac{g}{2 i} \phi^2 \psi^*,\\\label{phi}
i\partial_t\phi &=& J_c^2 G_{11}(\nu_a) \phi -\frac{g}{2 i} \psi^2 \phi^*.
\end{eqnarray}
For small $J_c$ we can neglect the real parts of $G_{11}$, as they just introduce a slight shift of the frequency. Note that $G_{11}(\nu_a)$ is complex and $\Gamma = -2 J_c^2 \rm{Im} G_{11}(\nu_a)$ represents  the inverse lifetime of the  antisymmetric mode.  This term must be kept in order to capture the physics.

If we throw away the real part of $G_{11}$, then both $\psi$ and $\phi$ are real. To explore the system's stability we  linearize about $\phi=0$, finding that $\phi$ decays to zero if $\Gamma>g\psi^2$.  Otherwise $\phi$ grows.  Thus there is a minimum amplitude $g$ needed to excite the system.  Such a threshold was seen in the experiment \cite{chin1}.

The number of particles in the symmetric mode, $\psi^2$, monotonically decreases with time. Thus if the drive is initially above the threshold ($g>\Gamma/\psi^2$), it will remain so for all time: Atoms will be ejected from the condensate until none remain. The emission rate $\Gamma \phi^2$ increases as the amplitude of the antisymmetric mode grows. At longer times, $\phi^2$ falls, and so does the emission rate. Thus the jet emission is in the form of a {\em pulse}.  For contrast, in our prior work the jet consisted of a relatively steady flux of particles.

\section{Numerics}\label{sec:numerics}

\subsection{Short-time behavior}
We verify the scenario from Sec.~\ref{sec:perturb} by numerically solving Eqs.~(\ref{a0})--(\ref{aj}). We assume that the system is in its equilibrium when $t<0$, and seed the antisymmetric mode by making $a_0$ and $b_0$ slightly different from one another: $a_0(t=0)=1.01$ and $b_0(t=0)=1$.
We work in units where $J_{ab}=1$, or equivalently measure energies and times in units of $J_{ab}$ and $\hbar/J_{ab}$.

\begin{figure}[tbp]
\includegraphics[width=\columnwidth]{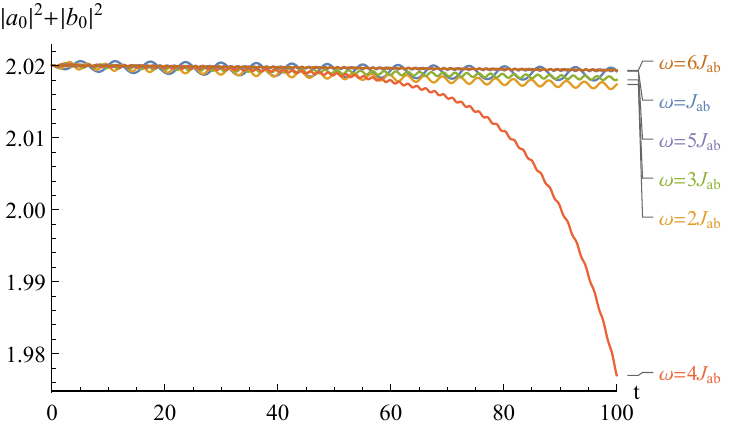}
\caption{(color online) Time dependence of the total number of trapped particles for different drive frequencies $\omega$. Here, the  potential depth is $V=-2$, and the drive strength is $g=0.1$. The coupling strengths are $J_c=0.1$ and $J_l=1$. Energies are in units of $J_{ab}$, and times are in units of $\hbar/J_{ab}$.}
\label{resofrecom}
\end{figure}

We first validate that significant particle emission only occurs when the drive frequency is tuned to resonance, $\omega=2(\nu_a-\nu_s)=4 J_{ab}$.  Figure~\ref{resofrecom} shows  the total number of trapped particles  $\vert a_0\vert^2+\vert b_0\vert^2$ as a function of time for different drive frequency $\omega$.  As can plainly be seen, the number of particles in the central sites is very stable, unless the drive is resonant.

\begin{figure}[tbp]
\includegraphics[width=\columnwidth]{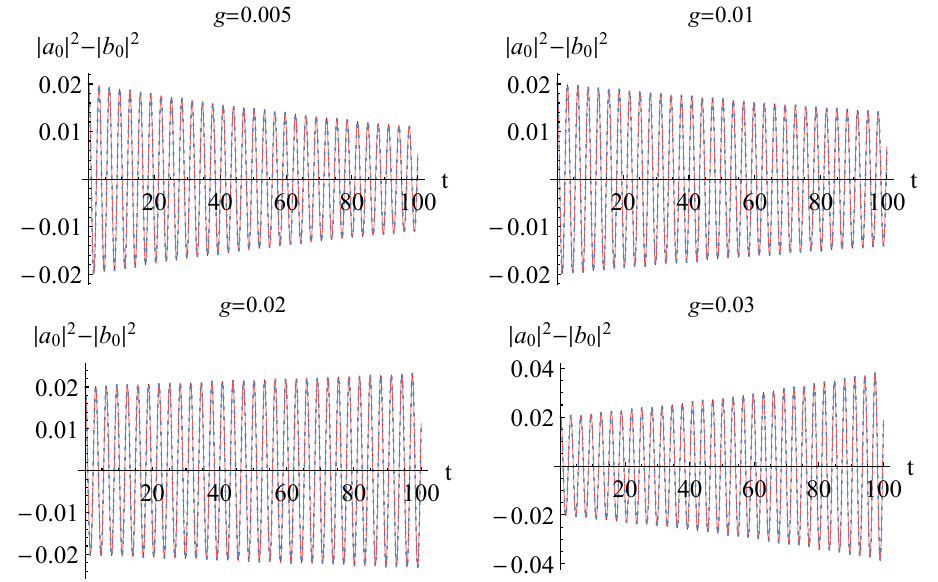}
\caption{(color online) Typical examples of the time evolution of the density modulation $|a_0|^2-|b_0|^2$ for different drive strengths $g$ with fixed trapping potential $V=-2$. The analytic results (red solid) and the numerical solutions (blue dotted) are nearly indistinguishable. The drive frequency is resonant, $\omega=4J_{ab}$, and the coupling strengths are $J_c=0.1$ and $J_l=1$. Energies are in units of $J_{ab}$, and times are in units of $\hbar/J_{ab}$.}
\label{aminusb}
\end{figure}

\begin{figure}[htbp]
\includegraphics[width=\columnwidth]{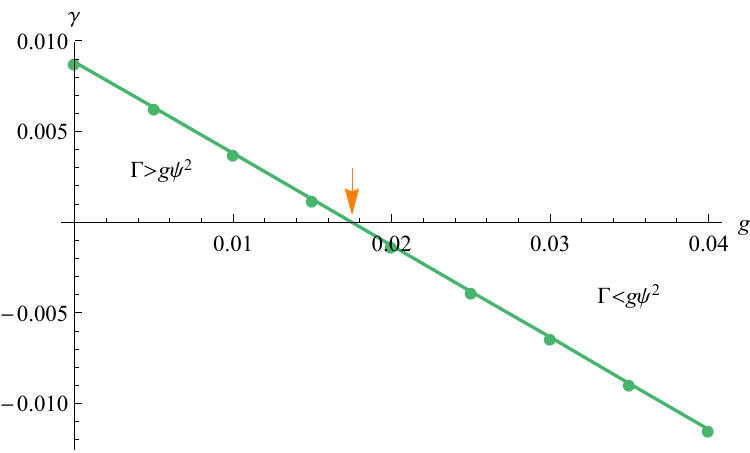}
\caption{Collective mode decay rate $\gamma$ found by fitting the envelopes of $|a_0|^2-|b_0|^2$ to a decaying exponential, up to time $t=100$. To the right of the arrow, $\gamma<0$, representing an exponential growth.   Here, we have taken $V=-2$, $J_c=0.1$ and $J_l=1$. The drive frequency is resonant. Energies are in units of $J_{ab}$, and times are in units of $\hbar/J_{ab}$.}
\label{threshold}
\end{figure}

Specializing to the resonant case, we next investigate the build-up of the antisymmetric mode. Figure~\ref{aminusb} shows the difference $|a_0|^2-|b_0|^2$ as a function of time for different drive strengths $g$. There is a clear separation of scales between the rapid oscillations and the slow  time-evolution of the envelope.  This separation of scales was key to the approximations in Sec.~\ref{sec:perturb}. As expected, when $g\psi^2<\Gamma$ the initial imbalance decays, while for larger drive the imbalance grows. In that figure we also plot the perturbative results from integrating Eqs.~(\ref{psi}) and (\ref{phi}).  The numerical and perturbative results are indistinguishable, and all figures appear to only have a single curve. To emphasize the threshold behavior, we fit the envelopes of these curves to exponentials, $(\vert a_0\vert^2-\vert b_0\vert^2)_{\rm env}=A e^{-\gamma t}$.  Figure~\ref{threshold} shows the exponential $\gamma$ as a function of the drive strength $g$, and compares it to the prediction from our perturbative analysis, $\gamma= \Gamma-g\psi^2$.

\begin{figure}[tbp]
\includegraphics[width=\columnwidth]{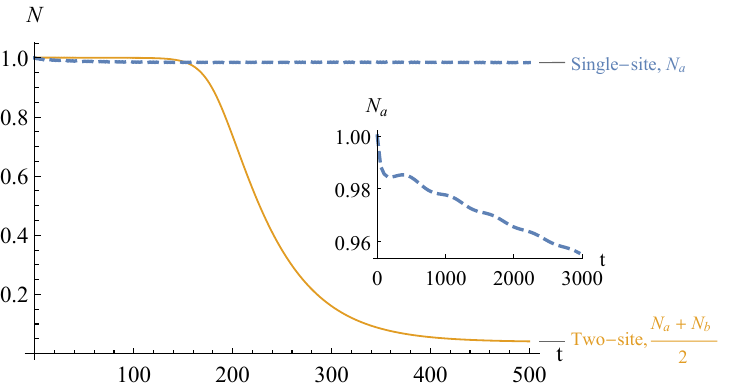}
\caption{
(color online)
Average number of particles per site in trap.  Orange solid: Two-site model described in this paper.  Blue dashed: Single-site model from Ref.~\cite{lai}. Here,
 $V=-2, g=0.1, J_c=0.1$ and $J_l=1$. In each case the  frequencies are tuned to resonance with $\omega_{\rm{S}}=2J_l$ and $\omega_{\rm{T}}=4J_{ab}$ for the single-site and the two-site, respectively.
 Inset shows the behavior of the single-site model in more detail, and over a longer time interval.
 Energies are in units of $J_{ab}$, and times are in units of $\hbar/J_{ab}$.
}
\label{decaycom}
\end{figure}

\begin{figure}[tbp]
\includegraphics[width=\columnwidth]{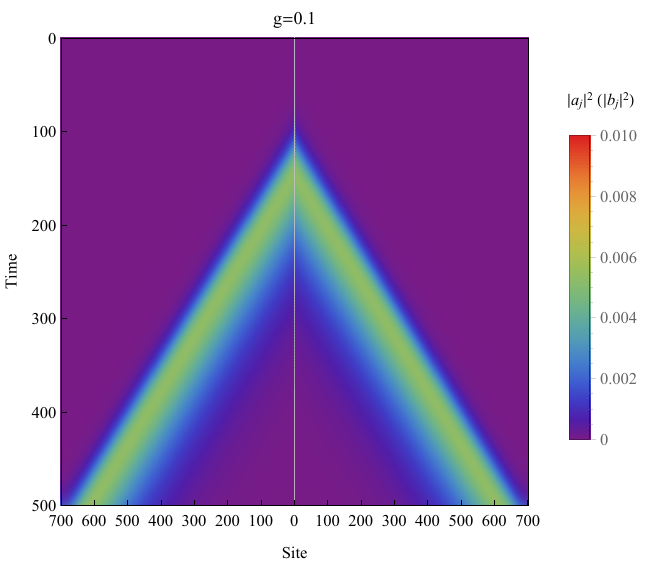}
\caption{
(color online) Number of the particles on the $j$th site of each lead, as a function of time. Here, $V=-2$,  $g=0.1$, $J_c=0.1$, and $J_l=1$. Energies are in units of $J_{ab}$, and times are in units of $\hbar/J_{ab}$.
}
\label{jetarray}
\end{figure}

\subsection{Long-time behavior}

Figures \ref{resofrecom}--\ref{threshold} illustrate the short-time behavior of the trapped particles.  In
Fig.~\ref{decaycom} we explore the full time-dependence of the average number of particles in each site of the trap.  For comparison, we have also included the results from the single-site model in Ref.~\cite{lai}.
We choose same values of $V,J_c,J_l$, and $g$ for both calculations, and use slightly different values of $\omega$, as the condition for particle emission is different in each case.  
The two-site model shows very little decay at short times, during which the antisymmetric mode grows in amplitude.  A large pulse of particles is emitted at intermediate times $200<t<300$, causing the number of trapped particles to rapidly fall.
The single-site case instead is characterized by a slow and steady decay, and the number of trapped particles is well approximated by an exponential. This behavior is best seen in the inset, which includes a longer time interval.
The collective mode in the two-site model provides a dramatic enhancement in the emission rate.

Finally, in Fig.~\ref{jetarray} we illustrate the structure of a jet by plotting the number of particles on every site of each lead, as a function of time.  The antipodal pulses are clearly visible.

\section{Summary and Outlook}\label{sec:summary}

We have produced the minimal model which can be used to explore how collective modes lead to a resonant enhancement of particle emission from a Bose-Einstein condensate with modulated interactions \cite{chin2,chin4}.  This model is designed so that much of its behavior can be analyzed analytically.  We validate our perturbative calculations through numerical studies.

In our model, the antisymmetric ``sloshing mode" is parametrically excited when the interactions are modulated at twice the collective-mode frequency (which in turn is the difference in energy between the antisymmetric and symmetric modes).  In this process, atoms are promoted to the antisymmetric state, where they can tunnel into the leads.  This tunneling  damps out the collective mode.  If the drive is stronger than the damping, the mode grows exponentially, leading to a large burst of particles.  Conversely, if the drive is weaker than the damping, then very few particles are emitted.

The underlying parametric resonance phenomena can be thought of as an amplifier:  when the drive is sufficiently strong, a small seed grows exponentially.  In our model there is only a single collective mode, so the end result is foreordained.  The experimental system, however, boasts a large number of collective modes.  Modulating the interactions will cause a number of these modes to grow exponentially, but one will dominate.   The final pattern will depend upon the initial fluctuations and the frequency of the drive \cite{chin4}.  The initial fluctuations are likely to be thermal in nature, in which case the dynamics are well described by the formalism used here.

It is interesting to contemplate the possibility that the parametric excitation could amplify quantum fluctuations.  In that case one would produce a quantum superposition of different collective modes, and a quantum superposition of different jets.
Such Schr\"{o}dinger cat states are quite fragile, and it would be challenging to detect the coherence between the macroscopically distinguishable configurations.  Nonetheless, the quantum nature of the fluctuations could be revealed in the statistics of the outcomes. There have been a number of relevant  optical analogs \cite{brooks,purdy}, and such parametrically driven systems have been proposed as platforms for quantum computing \cite{kerrcat}.

If one wanted to explore the physics related to the competition between different collective modes, one would need to extend our model to include more sites inside the trap.  One could introduce various seeds, and see how they grow, and study the properties of the resulting particle jets.  Geometries with more than two leads are particularly interesting.  Such configurations would enable a study of the correlation between the particle emission and the various leads.  The continuum limit of a large number of leads would mimic the experimental geometry, where antipodal particle jets are correlated.

\section*{Acknowledgments}
This work was supported by NSF Grant No. PHY-2110250. L.Q.L. received support from the China Scholarship Council (Grant No. 201906130092) and National Natural Science Foundation of China (Grant No. 11675051).




\end{document}